\documentclass[amsmath,amssymb,12pt,reprint,floatfix,unsortedaddress]{revtex4-1}

\usepackage{graphicx}
\usepackage{dcolumn}
\usepackage{bm}
\usepackage[utf8]{inputenc}
\usepackage[T1]{fontenc}
\usepackage{newtxmath,newtxtext}
\graphicspath{{plots/}{.}}
\usepackage{floatflt}
\usepackage{todonotes}
\usepackage{hyperref}
\usepackage[capitalize]{cleveref}
\usepackage{physics}
\usepackage{xspace}
\usepackage[caption=false]{subfig}
\usepackage[normalem]{ulem}
\usepackage[separate-uncertainty=true]{siunitx}

\newcommand{\Trun}{T_\mathrm{r}}
\newcommand{\Ttum}{T_\mathrm{t}}

\newcolumntype{Y}{>{\centering\arraybackslash}X}
\newcommand{\Us}{U_\mathrm{s}}

\newcommand{\uavg}{u_\mathrm{avg}}

\newcommand{\Koo}{K_{\mathrm{o,o}}}
\newcommand{\Kwo}{K_{\mathrm{w,o}}}

\newcommand{\ie}{{i.e.}}

\newcommand{\Ecoli}{{\it E.coli}}
\newcommand{\rhobo}{\overline{\rho^{\mathrm{behind}}_{\mathrm{obs}}}}
\newcommand{\rhowa}{\overline{\rho_{\mathrm{wall}}}}
\newcommand{\rhoo}{\rho_{\mathrm{obs}}}
\newcommand{\rhow}{\rho_{\mathrm{wall}}}
\newcommand{\rhoh}{\rho_{\mathrm{h}}}

\newcommand{\enterRearUp}{\mathcal{E}_{\mathrm{up}}}
\newcommand{\enterRear}{\mathcal{E}}
\newcommand{\Nup}{N_\mathrm{up}}

\newcommand{\ahem}[1]{``{#1}''}

\begin{document}

\title{The Influence of Motility on Bacterial Accumulation in a Microporous Channel}
\author{Miru Lee}
\email{miru.lee@uni-goettingen.de}
\affiliation{Institute for Theoretical Physics, Georg-August-Universit\"at G\"ottingen, 37073 
G\"ottingen, Germany}

\author{Christoph Lohrmann}
\email{clohrmann@icp.uni-stuttgart.de}
\affiliation{Institute for Computational Physics, University of Stuttgart, Allmandring 3, 70569 Stuttgart, 
Germany}

\author{Kai Szuttor}
\affiliation{Institute for Computational Physics, University of Stuttgart, Allmandring 3, 70569 Stuttgart, Germany}

\author{Harold Auradou}
\email{harold.auradou@universite-paris-saclay.fr}
\affiliation{Université Paris-Saclay, CNRS, FAST, 91405, 
Orsay, France}

\author{Christian Holm}
\email{holm@icp.uni-stuttgart.de}
\affiliation{Institute for Computational Physics, University of Stuttgart, Allmandring 3, 70569 Stuttgart, Germany}

\date{\today}

\begin{abstract}
  We study the transport of bacteria in a porous media modeled by
  a square channel containing one cylindrical obstacle via
  molecular dynamics simulations coupled to a lattice Boltzmann fluid.
  Our bacteria model is a rod-shaped rigid body which is propelled by
  a force-free mechanism. To account for the behavior of living bacteria, 
  the model also incorporates a run-and-tumble process.
  The model bacteria are capable of hydrodynamically interacting with both of 
  the channel walls and the obstacle.
  This enables the bacteria to get reoriented
  when experiencing a shear-flow. We demonstrate that this model is
  capable of reproducing the bacterial accumulation on the rear side
  of an obstacle, as has recently been experimentally observed
  by ~[G. L. Mi\~no, et al., Advances in Microbiology \textbf{8}, 451 (2018)]
  using \Ecoli{} bacteria. By systematically varying the
  external flow strength and the motility of the bacteria, we resolve
  the interplay between the local flow strength and the swimming
  characteristics that lead to the accumulation. Moreover, by changing
  the geometry of the channel, we also reveal the important role of
  the interactions between the bacteria and the confining walls for the accumulation
  process.

\end{abstract}

\maketitle

\section{Introduction}
\begin{figure*}
\centering
	\includegraphics[width=0.98\textwidth]{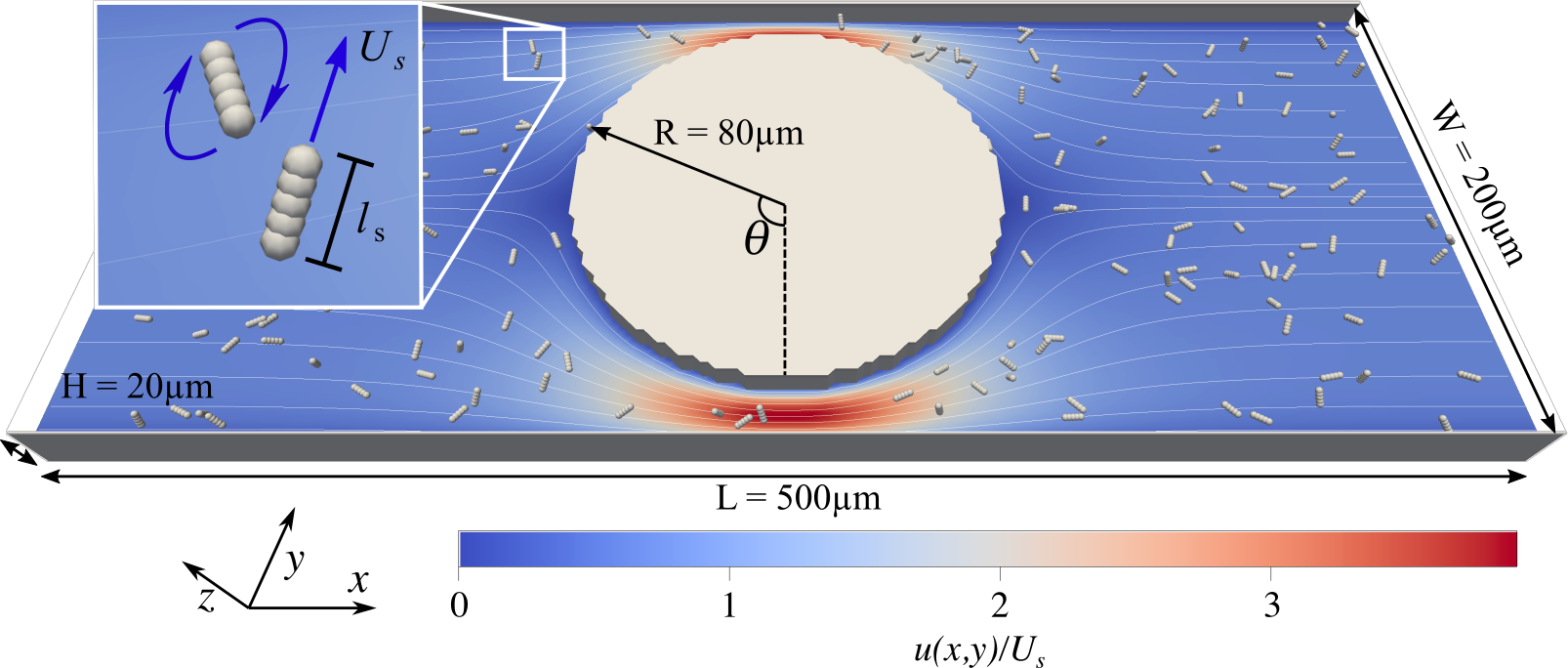}
	\caption{The simulation set-up consists of a rectangular
          fluid-filled channel of size
          $(L,W,H) =
          (\SI{500}{\micro\meter},\SI{200}{\micro\meter},\SI{20}{\micro\meter})$
          with a cylindrical obstacle of radius
          $\SI{80}{\micro\meter}$ placed at the center of the
          box. Inside the channel are up to 159 swimmers, each
          consisting of five interaction sites with the
          lattice-Boltzmann fluid, which are capable of performing
          run-and-tumble motions as indicated on the inset. The fluid
          is driven by an external force density. The stream lines and
          the colormap represent the driven flow-field projected on to
          the $xy$-plane normalized by the swimming speed $\Us$.}
	\label{fig:simul_box}
\end{figure*}
A growing number of emergent technologies take advantage of bacteria
metabolisms to provide novel, environmentally-friendly, and more
efficient alternatives to classical physicochemical methods.  This is,
for example, the case in the fields of pollution
reduction~\cite{pandey02a,aitken04a}, oil
recovery~\cite{donaldson89a,lazar07a,brown10a} biocalcification for
soil reinforcement ~\cite{dadda17a,marzin20a}, or to heal cement
cracks~\cite{demuynck10a,nimrat20a}.  However, one of the problems
researchers face when trying to optimize these processes is the
limited understanding of the role that bacterial motility plays at the
pore scale~\cite{alonso19a,yang19a,bhattacharjee19a,bhattacharjee19b,scheidweiler20a} and its coupling to local chemical gradients~\cite{deanna20a}.  Indeed, we now know
 that flagellated microorganisms such as \Ecoli{}, or sperm cells
display a great variety of swimming behaviors like upstream
motions~\cite{kaya12a,kantsler14a,figueroa15a}, drift trajectories on
surfaces~\cite{lauga06a,mathijssen19a} or helicoidal trajectories in a
flow~\cite{rusconi15a,junot19a}.  These motions contribute
significantly to the trapping of bacteria in pores as well as in
determining their localization in a channeled flow~\cite{rusconi14a} and their 
hydrodynamic dispersion~\cite{creppy19a,dehkharghani19a,scheidweiler20a}.
One of the consequences is an enhanced attachment and colonization of 
surfaces~\cite{rusconi14a,secchi20a,spagnolie15a,sipos15a} which influences the 
biofilm formation~\cite{drescher13a} and therefore the overall fluid 
flow~\cite{drescher13a,coyte17a}.
The aim of this work is to elucidate the important role that motility and
geometric confinement play at the pore scale for the accumulation of
bacteria on surfaces in a complex geometry.

Analytical tractable models and simulations can help to understand the experimental observations by elucidating the microscopic physics at play. The majority of the bacteria models incorporate mechanical torques by idealizing the shape of the bacteria, i.e., simple forms like rods or prolate particles~\cite{rusconi14a,ezhilan15a,ezhilan15b,secchi20a}. 
Despite the apparent simplicity of these models, they are
able to capture the complex helicoidal trajectories of microswimmers in a 
Poiseuille flow~\cite{rusconi14a,junot19a}, the accumulation of bacteria on 
flat surfaces~\cite{rusconi14a} or downstream obstacles~\cite{secchi20a}.
These models, however, exclude any particle-particle and particle-surface interactions, do not take into account the influence of the bacteria on the fluid flow, and neglect any steric effects.  
To overcome these limitations, we propose a hybrid modeling approach that couples the lattice-Boltzmann (LB) method to molecular dynamics (MD)~\cite{succi01a,kruger17a,duenweg09a,duenweg09b,ahlrichs99a,degraaf16a,degraaf16b}. Employing this method, we are able to reproduce trajectories of model bacteria in a porous medium in order to understand the mechanism of how bacteria interact with the surrounding fluid's movement and how they accumulate on surfaces.

Our model for a microporous environment consists
of a cylindrical obstacle placed under a microfluidic flow, mimicking
the experimental setup of Ref.~\cite{mino18a}.
This geometry has been chosen because it contains the basic ingredients found in porous media: solid surfaces, velocities that vary along the stream lines, and stagnant flow zones (regions of low velocities). 
The referred experimental article will also serve to validate our simulation 
approach. 
Our model for \Ecoli{} bacteria is based on the one proposed in Ref.~\cite{lee19a}. 
Here we couple hundreds of these dipolar force-free swimmers to a lattice-Boltzmann fluid to capture the hydrodynamic interactions between bacteria and a fluid, e.g., water.
We are thus able to follow single bacterial trajectories in detail, and by doing this for a sufficient number of trajectories we can study the bacterial distribution in our flow cell.

After having validated our simulation method, we will further
investigate the influence of the local flow speed on the accumulation
of the swimmers on the confining walls by varying the external flow
strength. This can be easily done as we have full control over the
flow boundary conditions. Since the change in confinement can alter
the swimming trajectory, we also compare a vertically open system
against the confined system. This sheds light on trajectories that
swimmers make when arriving at the obstacle, and thus on the role of
the bacterial interactions with the confining walls.  Next, the effect
of the run-and-tumble motion on the bacterial accumulation is
investigated, by changing the running duration scale from
almost passive Brownian particles to very persistent swimmers that
rarely change their swimming directions. We discuss the optimal
persistence in swimming that maximizes the accumulation.

The paper is organized as follows. In~\cref{sec:setup}, we explain the simulation details, i.e., the system geometry, the flow dynamics, and the swimmer model. 
\cref{sec:results} starts with demonstrating that our results are comparable with the existing experimental observation. We then further 
discuss the various factors that can affect bacterial accumulation. 
The paper ends with our concluding remarks in~\cref{sec:conc}.

\section{Simulation set-up}\label{sec:setup}
\subsection{Geometry and flow simulation}
The boundary geometry for the fluid and the bacteria is set up according to the experimental design in~\citet{mino18a} as shown in~\cref{fig:simul_box}. 
It comprises a rectangular channel of size $ L\cross W\cross H $ and a cylindrical obstacle of radius $R$ at the center of the box, which we define as the origin of our coordinate system. 
The frame of reference is the laboratory frame.

We solve the dynamic flow problem by means of the lattice-Boltzmann method~\cite{succi01a,kruger17a} which can be regarded as a Navier-Stokes solver.
The advantages of the algorithm are the simple implementation of complex boundary conditions and the possibility to couple the fluid simulation to the molecular dynamics simulation of the swimmers. 
Periodic boundary conditions are used along the $ x $-direction, and a no-slip boundary condition is imposed on the top and bottom surfaces (at $z=-H/2$ and $z=H/2$), on the lateral walls (at $y=-W/2$ and $y=W/2$) of the flow cell, and on the surface of the cylindrical obstacle.
The flow is driven through the channel by applying a constant force density onto each lattice-Boltzmann node. 
To test the correctness of our LB implementation and to investigate the severity of grid artifacts we performed a computational fluid dynamics simulation of the channel using a finite element method. 
The comparison can be found in the electronic supplementary information (ESI) Fig.~1~\cite{Note1}.

The system's geometry and the resulting flow field are depicted 
in~\cref{fig:simul_box}. We characterize the flow strength by the average value 
$ \uavg := \frac{1}{W\times 
H}\int_{-H/2}^{H/2}\int_{-\frac{W}{2}}^{\frac{W}{2}} u (x = 
\SI{250}{\micro\meter},y,z)\dd{y}\dd{z}$, i.e., measured at the outlet of the 
channel.
For all performed simulations, the Reynolds number of the flow is very small $\mathrm{Re}\sim 10^{-2}$. 
This is manifested through the spatial symmetries of the flow field with respect to the center of the box. 
Note that from now on, we always normalize the flow strength by the swimming 
speed $\Us$ of our model bacteria unless otherwise stated.

\subsection{Swimmer model}
In order to reproduce the elongated shape of \Ecoli~ the bacteria are modeled as a rigid collection of five aligned molecular dynamics particles (see~\cref{fig:simul_box}). 
A detailed description of the model is given in Ref.~\cite{lee19a}.

A short ranged non-bonded Weeks-Chandler-Anderson interaction
potential between the particles making up a swimmer is 
used to incorporate the swimmer's excluded volume. The body extension is  
$l_\mathrm{s}\sim\SI{5}{\micro\meter}$, and the flagella are not modeled 
explicitly. 
Thus, the aspect ratio (=body length/diameter) of $5$ for the swimmer is within the range of aspect ratios of particles used by \citeauthor{junot19a}~\cite{junot19a}. 

All molecular dynamics particles are coupled to the underlying lattice Boltzmann fluid via a point-friction coupling scheme according to Ref.~\cite{duenweg09b,ahlrichs99a,degraaf16a,degraaf16b}.

The swimmers perform a run-and-tumble motion as observed in some flagellated bacteria like \Ecoli{}. The dynamics are characterized by straight swimming phases (runs), interrupted by sudden changes in direction (tumbles)~\cite{berg72a,berg93a,saragosti12a}.

Straight swimming motion is, in the present study, obtained by applying a body-fixed force along the main axis of the swimmer. 
To ensure the force-free swimming mechanism and to mimic the propulsion by flagella rotation, a counter force of equal strength but opposite direction is applied to the fluid behind the swimmer~\cite{lee19a}.
During the runs, the swimmers can thus be understood as a force-dipole pusher with a constant swimming velocity $\Us$~\cite{degraaf16a,degraaf16b}.
The numerical parameters (listed in the ESI~\cite{Note1}) are chosen such that $\Us 
=\SI{24}{\micro\meter/\second}$, which is very close to the average velocity of 
the bacteria used by~\citet{mino18a}.

A tumbling motion is initiated by applying two opposite forces at the two terminating molecular dynamics particles of the swimmer, perpendicular to the swimmer's long axis. 
The two opposite forces create a rotating torque. Again, both forces are balanced by counter-forces on the fluid away from the swimmer to ensure a net-torque-free rotation.

During the simulation, the durations of run and tumble phases, as well as the tumble angles, are randomly drawn from the respective distributions. 
They reproduce the correct statistics of the run-and-tumble motion such that the (average) run and tumble durations are $\Trun = \SI{1}{\second}$ and $\Ttum=\SI{0.1}{\second}$, respectively~\cite{berg72a,berg93a,saragosti12a}. The swimmers' motion can thus be characterized by the run $\Trun$ and tumble $\Ttum$ durations as well as the swimming speed $\Us$.

Also note that rotation or translation of a swimmer through  thermal noise is not considered in our study since the effects of thermal diffusion are several orders of magnitude lower than those of the run-and-tumble motion.


We introduce $ N =159 $ swimmers to the system to achieve the low density of bacteria used in~\cite{mino18a}.
In the following we analyze the swimmer distribution in the channel in various situations. We hence define the swimmer distribution $\rho(\bm{r})$ as
\begin{equation}
    \rho(\bm{r})=\frac{1}{T}\sum_{i=1}^{N}\int_{0}^{T}\delta^3(\bm{r}_i(t)-\bm{r}) \dd{t},
    \label{Eq:density}
\end{equation}
where $r_i(t)$ is the $i$th swimmer's position at time $t$, and $T$ the total 
simulation length. It is a time-averaged one-particle distribution. The 
projection onto the $xy$-plane is then done by taking the average over the 
$z$-direction:
$\rho(x,y) = \frac{1}{H}\int_{-H/2}^{H/2}\rho(\bm{r})\dd{z}$.
We make the swimmer distribution $\rho(\bm{r})$ dimensionless by normalizing it with the homogeneous swimmer density $\rho_{\mathrm{h}}=N/V_\mathrm{box}$,
where $V_\mathrm{box}$ is the volume of the simulation box that is accessible to swimmers, i.e., excluding the volume occupied by the obstacle.

Here, we define some quantities that are useful in describing our observation. 
The swimmer distribution around the obstacle $\rhoo(\theta)$ as a function of 
polar angle $\theta$ is given by
$\rhoo(\theta)= \frac{1}{\SI{850}{\micro\meter^2}}\int_{\SI{80}{\micro\meter}}^{\SI{90}{\micro\meter}}\rho(r,\theta)r\dd{r}.$
Similarly, the swimmer distribution on the lateral walls $\rhow(x)$ at $y=-W/2$ 
and $y=W/2$ as a function of lateral position $x$ is
$\rhow(x)=\frac{1}{2\times\SI{10}{\micro\meter}}(\int_{-\SI{100}{\micro\meter}}^{-\SI{90}{\micro\meter}}\rho(x,y)\dd{y}+\int_{\SI{90}{\micro\meter}}^{\SI{100}{\micro\meter}}\rho(x,y)\dd{y})$. 
Consequently, we calculate the swimmer density behind the obstacle using 
$\rhobo = \frac{1}{\pi}\int_{\pi}^{2\pi} \rhoo(\theta)\dd{\theta}$ and on the 
lateral walls using $\rhowa = \frac{1}{L}\int_{-L/2}^{L/2} \rhow(x) \dd{x}$.

\section{Results}\label{sec:results}
\begin{figure*}
    \includegraphics[width=0.9\textwidth]{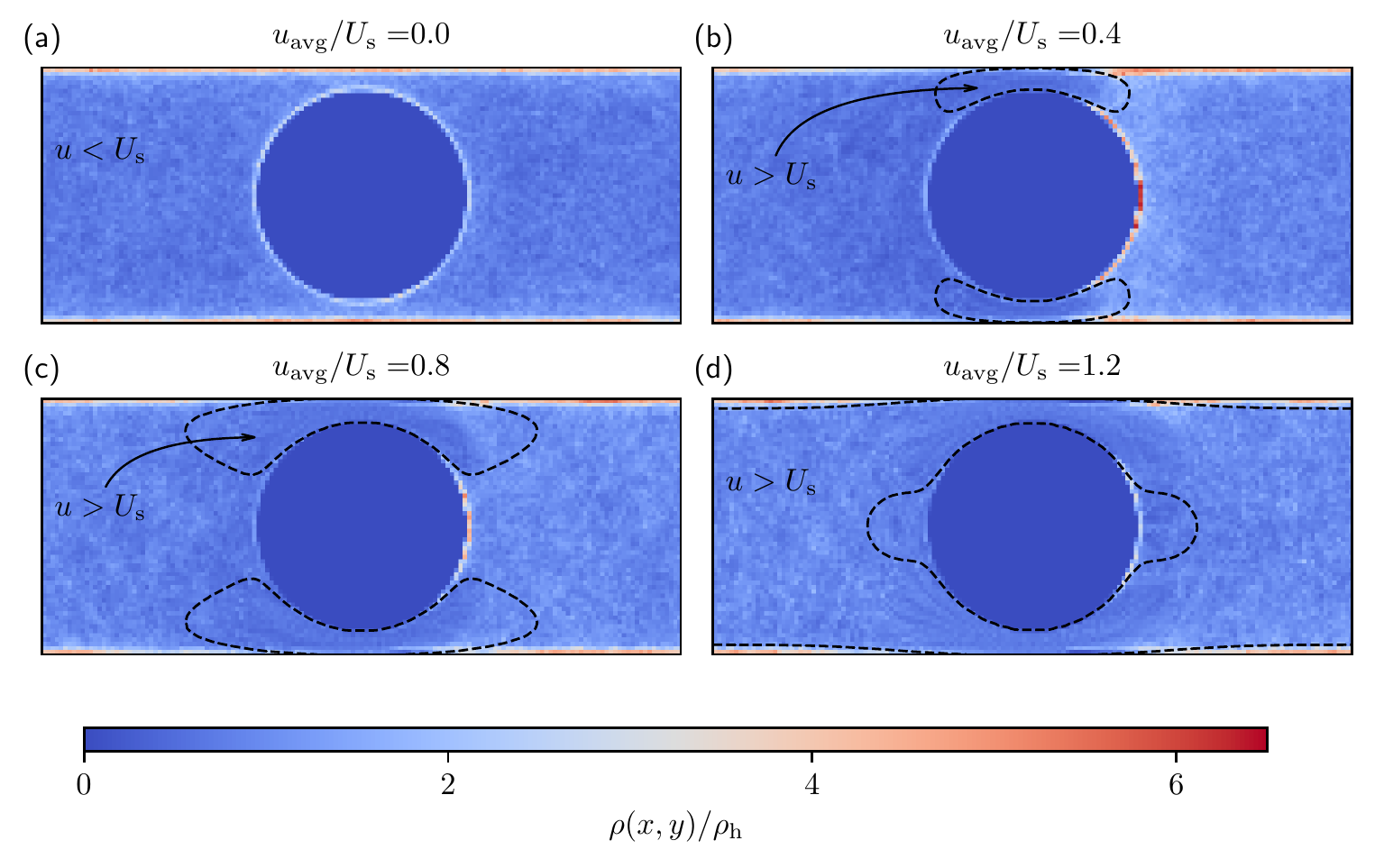}
   	\vspace{-0.5cm}
	\caption{The swimmer distribution $\rho(x,y)$ normalized with
          the homogeneous swimmer distribution $\rho_\mathrm{h}$ in
          the channel for various external flow inputs.  The dashed
          lines are contours, separating the regions where the
          magnitude of flow velocity $u(x,y)$, averaged in the
          $z$-direction, is greater than the magnitude of the swimming
          velocity $\Us$.  }
	\label{fig:dens_comp_rtm_dead}
\end{figure*}

\begin{figure}
	\centering
	\includegraphics{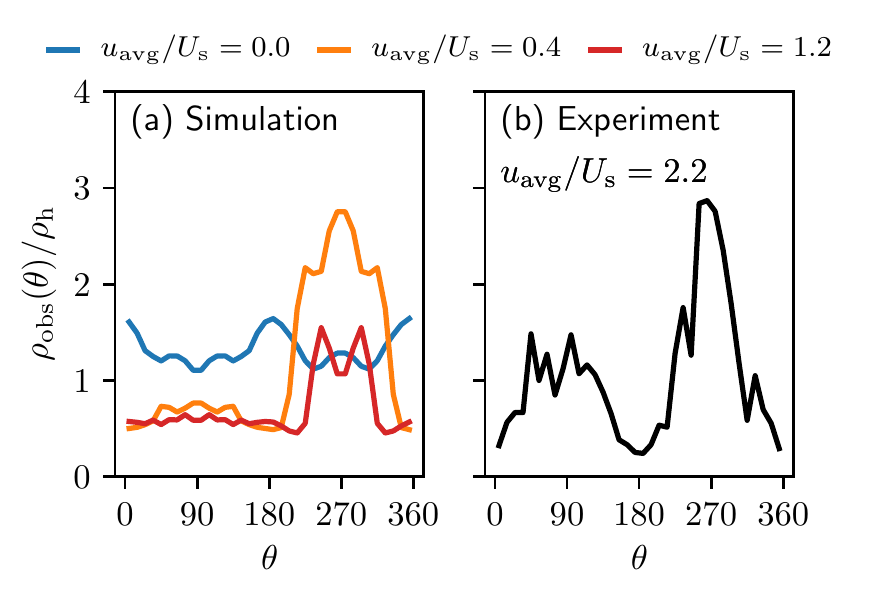}
	\vspace{-0.7cm}
	\caption{The normalized swimmer distribution around the
          obstacle $\rho_\mathrm{obs}/\rho_\mathrm{h}$ in (a) the
          simulation and (b) the experiment (Ref.~\cite{mino18a}) as a
          function of the polar angle around the center of the
          obstacle. $\theta=0$ and $180$ correspond to the lateral
          sides, and $\theta=90^\circ$ and $\theta=270^\circ$ to the
          upstream and downstream sides, respectively. The blue,
          orange and red lines in (a) stand for the velocity ratios
          $\uavg/\Us=0.0$, $0.4$, and $1.2$.}
	\label{fig:sim_vs_exp}
\end{figure}

\begin{figure}
	\centering
	\includegraphics{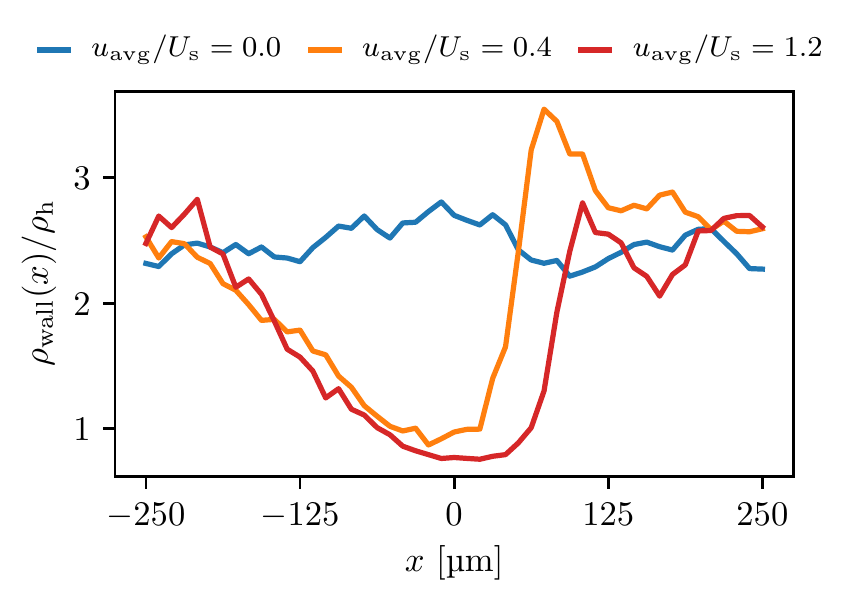}
	\vspace{-0.7cm}
	\caption{The normalized swimmer distribution along the lateral
          walls $\rho_\mathrm{w}/\rho_\mathrm{h}$ for different flow
          velocities.  The blue, orange and red lines stand for
          $\uavg/\Us=0.0$, $0.4$, and $1.2$, respectively.}
	\label{fig:wall_accu_hist}
\end{figure}

\begin{figure}
	\centering
	\includegraphics[width = 0.98\columnwidth]{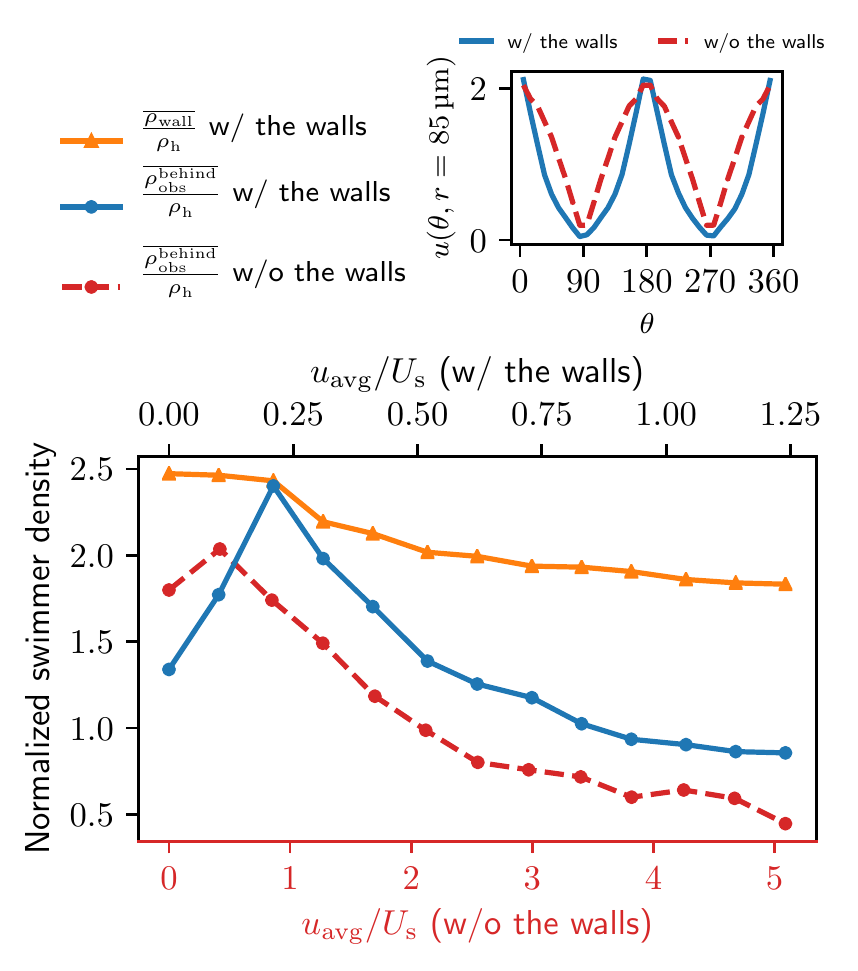}
	\caption{Main: The average swimmer density $\rhobo$ behind the
          obstacle (blue solid line), and $\rhowa$ on the walls
          (orange solid line) as a function of the external flow
          strength $\uavg/\Us$ in the presence of the lateral walls,
          as well as the average swimmer density behind the obstacle
          in the absence of the lateral walls (red dashed line). Top
          inset: the flow velocity $u(\theta,r=34\sigma)/\uavg$ around
          the obstacle as a function of polar angle $\theta$ for both
          cases, with (blue), and without (red) the lateral walls.}
	\label{fig:wall_vs_nowall}
\end{figure}

\begin{figure}
	\centering
	\includegraphics[width=0.98\columnwidth]{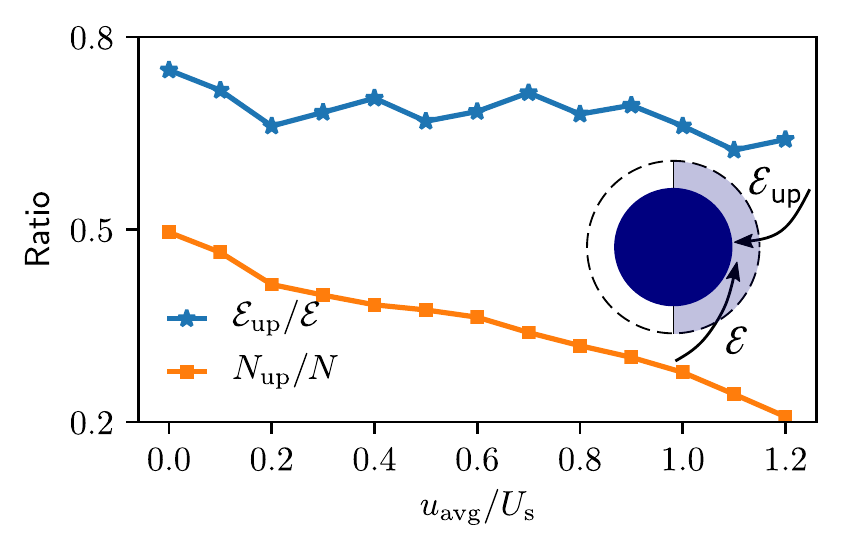}
	\vspace{-0.5cm}
	\caption{Blue solid line: The ratio of the number
          $\mathcal{E}_\mathrm{up}$ of events where upstream swimmers
          enter behind the obstacle to that of $\mathcal{E}$ events
          where swimmers enter behind the obstacle regardless of their
          swimming directions.  Red: The time averaged ratio of the
          number $\Nup$ of swimmers that swim upstream in the whole
          system to the total number $N$ of swimmers. A swimmer is
          counted as upstreaming if the $x$-component of its velocity
          in the laboratory frame is non-positive. Inset: A schematic
          visualization of events $\mathcal{E}$ and
          $\mathcal{E}_\mathrm{up}$.}
	\label{fig:upstream}
\end{figure}

\subsection{Accumulation is caused by niches}\label{subsec:niches}
\Cref{fig:dens_comp_rtm_dead} shows the spatial distribution of swimmers calculated using~\cref{Eq:density} inside the system box for different flow conditions. 
Additionally,~\cref{fig:sim_vs_exp}(a) and~\cref{fig:wall_accu_hist} 
quantitatively present the spatial distribution of the swimmers around the 
obstacle and on the lateral walls, respectively. 
Notice first that, in the absence of flow, a significantly large fraction of swimmers (i.e., $\rho > \rhoh$) is distributed both on the lateral walls and around the obstacle.
The bulges of the blue curve in~\cref{fig:sim_vs_exp}(a) at $\theta=0^\circ$ and $180^\circ$, as well as that of the blue curve in~\cref{fig:wall_accu_hist} at $x=0$ show a small enhancement of the accumulation at places where the obstacle and the lateral walls are closest. We refer to these regions as constrictions.
The homogeneous swimmer accumulation on the surfaces, i.e., both on the lateral walls and on the obstacle, is to be expected~\cite{elgeti13a,elgeti15a,volpe11a,zeitz17a}, since at the chosen running duration, the persistent swimming length scale ($l_\mathrm{per}=\SI{24}{\micro\meter}$) is comparable to the length scales in the channel; 
because the swimmers swim in all possible directions with equal probability, at some point they will touch a surface and stay there until tumble events orient their swimming directions away from the surface.

Additionally, we observe that more swimmers are accumulated on the lateral walls than around the obstacle. This can be explained by the geometric characteristics of the surfaces. The obstacle is a convex surface, and not capable
of containing swimmers for as long as the flat walls can 
do~\cite{spagnolie15a,sipos15a}. 
We attribute this to the simple fact that the swimmers will depart from 
any convex surface merely by swimming straight in any direction that was initially tangent to the surface.
The influence of the convexity will become more significant with increasing running duration $\Trun$.

Introducing an external flow, we measure an inhomogeneous distribution of swimmers on the surfaces. 
A larger number of swimmers accumulates on the downstream side of the obstacle while the density of swimmers on the upstream side of the obstacle is reduced, falling below the homogeneous swimmer density $\rhoh$. 
For $\uavg/\Us = 0.2$  nearly three times more swimmers per unit of volume are located at the rear of the obstacle than anywhere else in the fluid. 
This finding is in agreement with the observation made in the experiment~\citet{mino18a}.

We furthermore find that stronger flow velocities reduce the extension of the regions where the accumulation is observed. 
Consequently, the swimmer densities $\rhobo$ behind the obstacle and $\rhowa$ on the lateral walls reduce with increasing external flow strength, as indicated by the solid lines in~\cref{fig:wall_vs_nowall}. 
To explain this, we mark the regions where the magnitude of the local flow velocity is higher than the swimming velocity $\Us$ in~\cref{fig:dens_comp_rtm_dead}. 
For $\uavg/\Us<1.0$, the regions  where $u>\Us$ are localized in the constrictions. 
At the strongest external flow ($\uavg/\Us = 1.2$), the region covers the entire channel apart from two small domains located at the rear and front of the obstacle, and parts of the lateral walls located away from the constrictions.

Higher local flow speed regions act as one way streets; all swimmers are moving down-stream regardless of their swimming direction, since they cannot compete with the flow. 
The borders of the stronger flow regions therefore act as a \ahem{niches} in which the upstream swimmers stay until they reorient. 
This effect leads to an asymmetric distribution of swimmers, with a higher density in the right half of the channel.

The niche argument implies that many swimmers that accumulate behind the obstacle are swimming against the local flow direction. 
To support this argument we calculate, as shown in~\cref{fig:upstream},
the ratio of the number $\enterRearUp$ of events where a swimmer enters the accumulation region by swimming upstream to the total number $\enterRear$ of entering events. 
This ratio stays roughly constant at a high value of about $70\%$ irrespective 
of the increasing external flow.
This is in contrast to the ratio of upstream swimming bacteria in the whole system ($\Nup/N$), which decreases monotonically with increasing external flow.
From the two curves we can conclude that the smaller number of accumulated swimmers behind the obstacle is due to the fact that the total number of swimmers that are capable of accumulating is reduced.
The mechanism of accumulation itself (upstream swimming into niches) remains unaltered despite the increasing external flow.

The accumulation behind the obstacle (blue curve in \cref{fig:wall_vs_nowall}) displays a non-monotonic behavior which can be explained as follows. 
With a very weak external flow, the available space for the swimmers to accumulate is relatively large. 
The number of swimmers reaching the rear is thus reduced, because a large fraction is accumulated elsewhere.
The accumulation exhibits a maximum around $ \uavg/\Us = 0.2 $. 
This value coincides with the flow strength at which the local flow speed at the constriction becomes larger than the bacterial swimming speed. 
The one way street mechanism now leads to the maximum accumulation because the constrictions effectively block the upstream swimmers but the overall flow speed is not yet strong enough to flush the swimmers.
With a further increase of the external flow strength, the size of the niche shrinks, as can be observed in \cref{fig:dens_comp_rtm_dead}. 
Naturally, this limits the accessible surface area, and therefore the accumulation decreases.

\subsection{Role of lateral walls on accumulation}
\begin{figure}
  \centering \includegraphics[width =
  0.98\columnwidth]{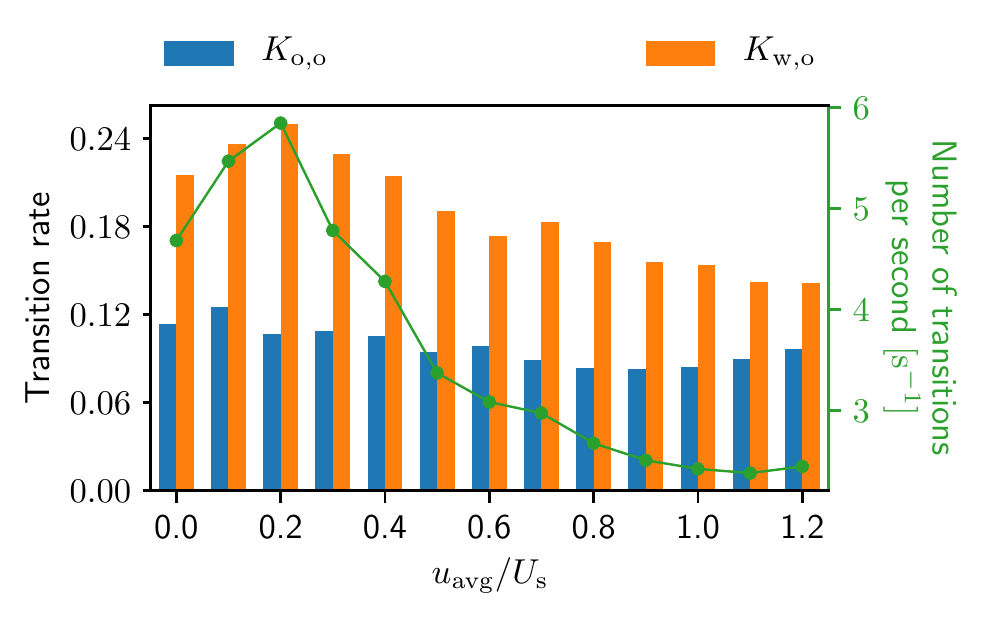}
  \caption{Transition rates $K_{\mathrm{o,o}}$ for the
    from-obstacle-to-obstacle and $K_{\mathrm{w,o}}$ for the
    from-wall-to-obstacle as a function of external flow strength. The
    green curve represents the combined number of transitions per
second. More detailed information can be found in Fig.~2, ESI~\cite{Note1}.}
	\label{fig:transition_uavg}
\end{figure}
In this part we will analyze in detail the effects that the lateral walls play in the bacterial accumulation.
Looking back again into \Cref{fig:dens_comp_rtm_dead}, we also note the preferential accumulation of the bacteria in the right half of the channel. 
This means that the niche argument also applies to the accumulation on the lateral walls. Moreover, a significant number of bacteria accumulate on the walls regardless of external flow strength. 
We argue that these accumulated swimmers on the walls can potentially migrate to the obstacle.

Due to the geometry, the lateral walls orient the swimmers to the $\pm x$ directions as they slide along. 
The upstreaming fraction can travel along the channel even under strong external flow, as the no-slip boundary condition provides niches of low flow velocities. 
Using this route, a swimmer can move up to the constriction with a high probability. 
Behind the constriction, the streamlines fan out and depart from the wall.
This flow away from the walls causes the bacteria to reorient and turn towards to the cylinder, as can be seen in Vid.~S1, ESI\footnote{\label{footnote_ESI}Electronic supplementary information (ESI) available. See DOI: xx.xxxx/xxxxxxxxxx}.

The swimmers' transitions from one surface to another can be easily calculated from the trajectories, and the results are displayed in~\cref{fig:transition_uavg}.  
In~\cref{fig:transition_uavg} one can see that the from-wall-to-obstacle transition $\Kwo$ happens more often than the from-obstacle-to-obstacle transition $\Koo$ across the board.
Thus, without lateral walls, one may expect a smaller accumulation around the obstacle.

To quantify our argument above, we also performed a new set of simulations in which the lateral walls were removed and replaced by a periodic boundary condition in the $y$ direction. 
To make the new system as comparable to the one with walls, the system size is 
also changed to $(L,W,H) = 
(\SI{500}{\micro\meter},\SI{500}{\micro\meter},\SI{20}{\micro\meter}).$ 
In addition, we adjusted the force densities, such that the flow velocity and profile around the obstacle are as close as possible to the original geometry (see the top right inset in~\cref{fig:wall_vs_nowall}). 
The total number of swimmers is changed from 159 to 248 to obtain better statistics.
Notice that the change in the total number of swimmers does not affect the overall dynamics of the total system since we remain in the low density limit.

In the absence of lateral walls, only a relatively small accumulation behind 
the obstacle is found, except for the first two smallest external flow 
conditions as represented by the red dashed line in~\cref{fig:wall_vs_nowall}.
This is because with a very weak external flow strength, the swimmers can accumulate on any surface. 
Without the lateral walls, it is thus natural that more swimmers accumulate at 
the 
cylinder.
With a stronger external flow strength, however, the swimmers without a lateral 
confinement will only end up behind the obstacle if a tumble happens at the 
right time with the right angle to allow them to come close to the surface of 
the obstacle. 
Therefore, for most of the time, the bacteria are just following the fluid flow.

Notice also that in~\cref{fig:transition_uavg} the green curve, \ie, the total number of transitions per second, as well as the from-wall-to-obstacle transition $\Kwo$ as a function of the external flow strength resemble that of the normalized swimmer density in~\cref{fig:wall_vs_nowall}. 
To explain this we resort to~\cref{fig:wall_accu_hist}. 
If the external flow strength is very weak, the swimmers on the walls are distributed rather homogeneously, meaning that many of the swimmers are located far away from the obstacle. 
The migration from the walls to the obstacle, consequently, happens less frequently. 
As the external flow gets stronger, however, the swimmers preferably accumulate on the wall at $x\sim\SI{120}{\micro\meter}$, which is close to the obstacle. 
Now the swimmers have to overcome only a shorter distance to arrive at the 
obstacle, which enhances the from-wall-to-obstacle transition $\Kwo$.

\subsection{Influence of swimming characteristics}
\begin{figure}
	\centering
	\includegraphics[width = 0.98\columnwidth]{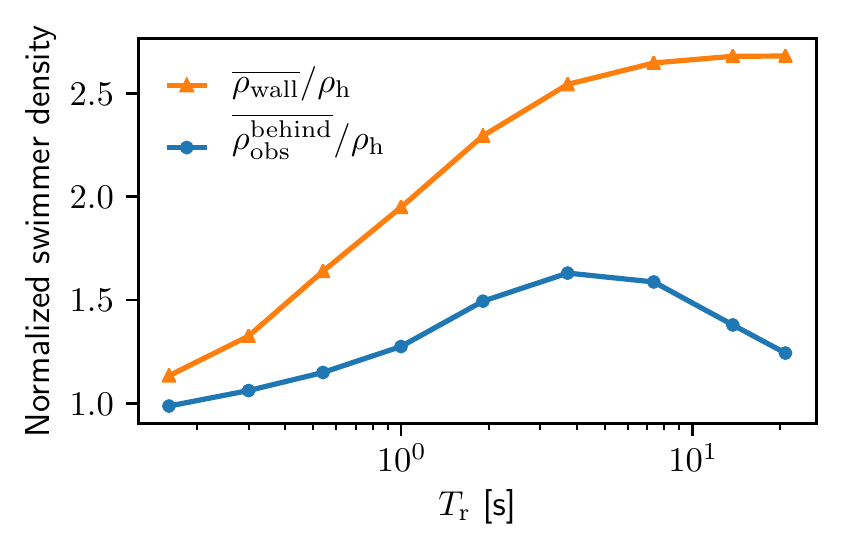}
	\caption{The normalized swimmer density
          $\rhobo/\rho_\mathrm{h}$ behind the obstacle (blue solid
          line) and $\rhowa/\rho_\mathrm{h}$ on the lateral walls
          (orange solid line) as a function of the (average) running duration $\Trun$}
	\label{fig:trun_accu_size}
\end{figure}

\begin{figure}
	\centering
	\includegraphics[width = 0.98\columnwidth]{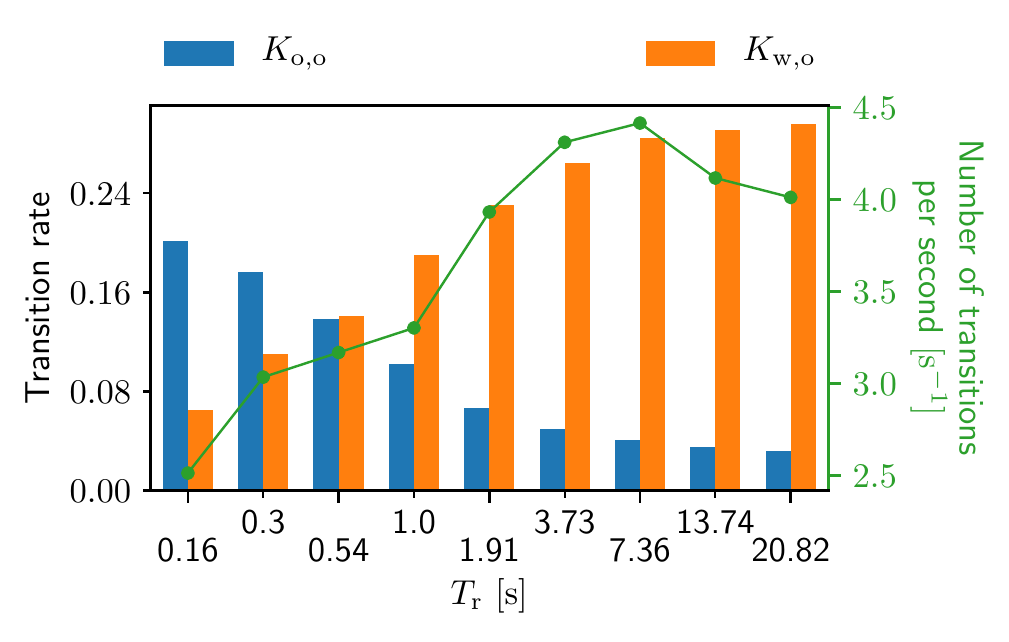}
	\caption{Transition rates for the from-obstacle-to-obstacle
          $K_{\mathrm{o,o}}$ and for the from-wall-to-obstacle rate
          $K_{\mathrm{w,o}}$ as a function of the (average) running duration $\Trun$. The green curve represents the combined
          number of transitions per second. More detailed information can be found in Fig.~3, ESI~\cite{Note1}.}
	\label{fig:transition_trun}
\end{figure}
In order to understand the physics behind the bacterial accumulation mechanisms better, we also investigate the influence of the running duration $\Trun$. We keep the external flow strength fixed at $\uavg=0.6\Us$.

In \cref{fig:trun_accu_size} we display the swimmer densities behind the 
obstacle and on the lateral walls as a function of $\Trun\in[\SI{0.16}{\second},\SI{21}{\second}]$. 
Intriguingly, the swimmer accumulation behind the obstacle peaks, and then decreases, whereas the bacterial density on the walls monotonically increases.

When $\Trun$ is very small, the behavior of the swimmers is similar to Brownian motion of passive particles.
As they change their direction rapidly, their swimming only leads to enhanced diffusion, but not to persistent motion.
For a visualization of this effect we refer to Vid.~2, ESI~\cite{Note1}.
The lack of persistent motion yields a very small accumulation density on the boundaries.

As $\Trun$ increases, the swimmers start showing an increasing  persistent and directed motility that allows them to swim for a sufficient amount of time to reach the boundaries.

With a very large $\Trun(>\SI{7}{\second})$, however, the situations on the walls and behind the obstacle start diverging.
This is primarily due to the shape of the boundaries as mentioned in~\cref{subsec:niches}.
The swimmers with very high $\Trun$ rarely change their directions. 
Therefore, the walls can trap swimmers much longer than the obstacle. Note that 
similar observations are reported by 
\citeauthor{spagnolie15a}~\cite{spagnolie15a} and 
\citeauthor{sipos15a}~\cite{sipos15a}. Although without flow, the studies 
identify an optimal obstacle size for a hydrodynamic capture, and point out, 
as in the present study, the key role of a surface geometry on the accumulation 
of microswimmers.

The running duration plays a substantial role in the transition behavior as well (see~\cref{fig:transition_trun}). 
As $\Trun$ gets larger, the migration from the walls to the obstacle happens more often. 
On the other hand, the from-obstacle-to-obstacle transition happens less frequent with increasing $\Trun$. 
One shall keep in mind, however, that this transition at a very small $\Trun$ is a rather trivial transition. At very small $\Trun$ a swimmer is essentially a passive Brownian particle, jiggering back and  forth to the obstacle, making a number of meaningless \ahem{transitions}. 
Such  transitions, however, become less probable as $\Trun$ increases  (also see Vids.~1 and 3, ESI~\cite{Note1} and compare the swimmers' behaviors).

The number of transitions per second in~\cref{fig:transition_trun} also increase with growing $\Trun$ until $\Trun = \SI{7.36}{\second}$, and then start decreasing. 
Fixed by the system's geometry, we  can find the optimal running duration for the accumulation behind the obstacle around $\Trun \sim 4$ from~\cref{fig:trun_accu_size,fig:transition_trun}.
It is worth noting that the swimming P\'eclet number, the ratio of the 
persistent running length $(=\Us\Trun)$ to the body size, is not a control 
parameter when it comes to the bacterial accumulation in porous media. This is 
because the ratio of local fluid flow speed to the swimming speed also affects 
the accumulation as discussed above.

\subsection{Limitations of the coarse-grained bacterial model}

Our bacterial model and the simulation could reproduce qualitatively the preferred accumulation behind the obstacle as observed in the experiment of~\citet{mino18a}, but there remains a quantitative discrepancy. 
The simulations overall yielded a smaller accumulation density around the obstacle than found in the experiment. 
In the simulation, the swimmers were mostly washed away when the average flow speed exceeded $\uavg=1.2\Us$, whereas in the experiment, the bacteria could manage to accumulate even under a stronger external flow of $\uavg\sim2.2\Us$. 
The difference in accumulation density was found particularly pronounced in front of the obstacle.

Despite the fact that one should not over-interpret results obtained by coarse-grained models, we present three reasons for this discrepancy. 
First, the swimming speed of our bacterial model was kept constant in the simulation in order to achieve a better understanding of the interplay between the local flow field and the swimmers' motility. 
In the experiment, the bacterial swimming speed distribution follows a half-normal distribution with a standard
deviation that is roughly of size ~$\Us$~\cite{mino18a}. 
This means that numerous
bacteria are able to swim faster than~$\Us$ and therefore more of them can accumulate behind the obstacle.
In addition to this, the discrepancy may also be due to the fact that we did not introduce 
any attractive interaction between the swimmers and the boundaries. 
It is well known that \Ecoli{} can adhere to surfaces via a short range electrostatic interaction~\cite{li04b, ong99a}, and they also can interact with the obstacle via pili.
Finally, we neglected the rotation of the bacterial body around its main axis and the counter-rotation of the flagella that causes the bacteria to swim in circular trajectories on surfaces~\cite{lauga06a}.
This will result in a lower effective diffusivity, and can cause the bacteria to explore less space in a given time compared to straight swimming.
In front of the obstacle, bacteria will escape the region of small flow by swimming in any direction (except straight into the cylinder), so only circular swimming could cause the prolonged residence time in this area.

\section{Conclusions}\label{sec:conc}
Our simulations demonstrated that
motile microorganisms preferably accumulate in regions where the fluid
speed is lower than the swimming speed, which are referred to as niches. For the geometry considered
here, the niches are located behind the obstacle (in direction of the
external force density) and on the lateral surfaces. Especially, when it comes to the accumulation behind the obstacle, we showed that upstream swimming of swimmers plays an important role.
This conclusion is in line with the recent results reported by 
\citeauthor{alonso19a}~\cite{alonso19a} and 
\citeauthor{secchi20a}~\cite{secchi20a}. In the first study, they investigated 
the dispersion of swimmers in a matrix of an obstacle, whose shape is 
systematically altered from a circle to an ellipsoid and to a triangle. They 
showed that, as long as an external flow is moderate, such an upstream swimming 
pattern can be observed not only with circular obstacles but also with 
triangular obstacles, the edges of which are pointing to the downstream 
direction. This suggests that the niche argument is not restricted to a 
cylindrical obstacle.
In the second study, they used a microfluidic chip containing circular pillars of different diameters, and observed more bacteria at the downstream side of the obstacles. 
They identified the shear induced reorientation as the mechanism allowing the bacteria to accumulate in these specific regions. Using a similar approach, \citeauthor{slomka20a}~\cite{slomka20a} came to the conclusion that the reorientation is also responsible for the accumulation of motile bacteria at the rear of a sinking spherical particle. In both studies, the shear induced by the flow around the obstacles is the physical mechanism, explaining the accumulation of the bacteria.
In the present study, the confinement by lateral walls also
plays a substantial role on the swimmer accumulation behind the
obstacle.  These walls produce additional zones of small fluid
velocity due to the no-slip boundary condition which provide a pathway
for the bacteria to swim upstream.  This mechanism is an important
element since it allows swimmers to come closer to the constrictions,
from which the swimmers can migrate to the obstacle. 
The accumulation of bacteria by the surface is triggered by the local shear that reorients the bacteria toward the surface. We note that a model where bacteria are replaced by rod like particles~\cite{rusconi14a,secchi20a} that are reoriented by the shear is sufficient to capture this process. However, because the hydrodynamic and the steric interactions between the surface and the bacteria are the key mechanisms leading to the "trapping" of the bacteria by the surface~\cite{lauga07a}, rod-like approaches fail to model the swimming along the surface. Our model includes the hydrodynamic and steric effects, and thus is successful in taking into account the effect of the lateral surfaces on the accumulation.
Finally, we observe that an optimal bacterial accumulation can be achieved when the running duration is around \SI{7}{\second}. These observations can help to design and optimize
strategies to sort and trap microorganisms.
It also can reveal insights into the physical mechanisms important for the filtration of motile bacteria in porous media.\\

Our study shows that the bacterial model coupled to a LB fluid
is an efficient technique to simulate motile
microorganisms at the pore scale including hydrodynamic and steric interactions.
The LB method enables us, in principle, to consider arbitrary complex
3D pore geometries and flow conditions~\cite{yiotis13a} easily. For these reasons, we believe that our approach can be used in future work to study the influence of flow on the "hopping and trapping" complex dynamics of bacteria recently observed in confined environment~\cite{bhattacharjee19a,bhattacharjee19b}. A strong advantage of the model is the ability to simulate many
interacting micro swimmers, opening up the possibility to study
collective effects in denser solutions.  The bacterial swimming
characteristics can be easily adapted to reproduce other types of
bacteria.  In the present article, the swimmers are of the pusher
type, like \Ecoli{}, but the method can be implemented to consider
neutral swimmers or pullers as well.  Some algae, for example, fall
into the last category.  Another advantage of our model is that the
persistence swimming time and the tumbling dynamics can also be
modified to incorporate more complex stochastic dynamics.
In total our model could be easily extended to investigate other important systems  from the point of view of applications.\\

\section*{Acknowledgments}
M.L, C.L., K.S. and C.H. were funded by the Deutsche
Forschungsgemeinschaft (DFG, German Research Foundation) – Project
Number 327154368 – SFB 1313, and the SPP 1726 ``Microswimmers: from
single particle motion to collective behavior'', grant HO1108/24-2.
H.~A. is supported by public grants overseen by the French National
Research Agency (ANR): (i) ANR Bacflow AAPG 2015 and (ii) from the
“Laboratoire d’Excellence Physics Atom Light Mater” (LabEx PALM) as
part of the “Investissements d’Avenir” program (ref- erence:
ANR-10-LABX-0039-PALM).

\bibliography{bibtex/icp}
\bibliographystyle{unsrtnat}
\end{document}